\documentclass[conference,a4paper]{IEEEtran}
\usepackage[T1]{fontenc}% optional T1 font encoding
\usepackage{cite}
\usepackage{url}
\usepackage{amsmath}
\usepackage{amssymb}
\usepackage[cmintegrals]{newtxmath}
\usepackage{subcaption}
\usepackage{dblfloatfix}
\usepackage{graphicx}
\usepackage{epstopdf}
\usepackage{xcolor}
\usepackage{mathtools}
\usepackage[detect-weight=true, binary-units=true]{siunitx}
\usepackage[inline]{enumitem}
\usepackage{lettrine}
\usepackage[boxruled,norelsize]{algorithm2e}

\makeatletter
\newcommand{\removelatexerror}{\let\@latex@error\@gobble}
\makeatother

\begin{document}
\title{Optimal Resource Dedication in Grouped Random Access for Massive Machine-Type Communications}

\author{\IEEEauthorblockN{Bin~Han, Mohammad~Asif~Habibi and Hans~D.~Schotten}
\IEEEauthorblockA{University of Kaiserslautern\\
Institute for Wireless Communication and Navigation\\
Paul-Ehrlich-Stra\ss e 11, 67663 Kaiserslautern, Germany\\
Email: \{binhan, asif, schotten\}@eit.uni-kl.de}
}

\IEEEoverridecommandlockouts
\IEEEpubid{\makebox[\columnwidth]{\copyright~Copyright 2017
		IEEE \hfill} \hspace{\columnsep}\makebox[\columnwidth]{ }}

\maketitle

\begin{abstract}
The high risk of random access collisions leads to huge challenge for the deployment massive Machine-Type Communications (mMTC), which cannot be sufficiently overcome by current solutions in LTE/LTE-A networks such as the extended access barring (EAB) scheme. The recently studied approaches of grouped random access have shown a great potential in simultaneously reducing the collision rate and the power consumption in mMTC applications, and exhibit a good compatibility with the concept of random access resource separation. In this work, we propose an optimized resource dedication strategy for grouped random access approaches, which inherits the advantage of resource separation to isolate device classes from each other, while providing an optional class preference with high flexibility and accuracy, which has been usually implemented with access class barring.
\end{abstract}


% \fi
\IEEEpeerreviewmaketitle

\section{Introduction}
%For conference papers, commonly not to make the 1st letter larger
The concept of Machine-Type Communications (MTC) refers to automated communication services and applications between devices or machines without human intervention. Driven by the explosive growth of demand in Internet-of-Things (IoT), future mobile networks of the $5^\textrm{th}$ Generation (5G) are expected to serve a massive number of MTC devices (MTCDs). The call for supporting massive MTC (mMTC) leads to various technical challenges, including the Radio Access Network (RAN) congestion as one of the most important issues. 

RAN congestions are caused by Random Access (RA) collisions, which occur when multiple devices simultaneously attempt to access the network with a same RA preamble. Due to the huge amount of MTCDs and the common synchronization among MTCDs, mMTC is supposed to easily create dense RA collisions, and to significantly increase the risk of RAN congestion. This is hardly to be solved in the framework of legacy cellular systems, such as Long Term Evolution (LTE) and LTE-Advanced (LTE-A) networks, whose radio access domain is designed to deal with a low number of connections. New mechanisms are therefore under investigation to fulfill the requirements for emerging networks.  

Various approaches have been proposed to reduce RA collisions in LTE/LTE-A networks, as deeply reviewed in \cite{ali2017lte}. Among them, six solutions have been specified by the \textit{$3^\textrm{rd}$ Generation Partnership Project (3GPP)}, including {Access Class Barring (ACB)}, {Backoff}, {Dynamic Resouce Allocation}, {Slotted RA}, {RA Resource Separation} and {Pull-Based RA}\cite{3gpp2011study}. Especially, ACB has been integrated into LTE-A since Release 10, and further developed to the so-called extended access barring (EAB) scheme \cite{3gpp2011study}.

\IEEEpubidadjcol 
Generally in ACB approachs, a set of access classes (ACs) are provided to classify devices, and correspondingly a set of barring time durations. The eNodeB (eNB) determines an access probability $p$ depending on the current RA collision density and broadcasts it in the local cell. Each device generates its own access probability $q$ w.r.t. its AC, and compare it to $p$. An access request is sent if $q\le p$, otherwise the device waits for a barring time duration before the next attempt. Through this mechanism, the network can not only flexibly compromise the access delay with the collision rate, but also separately configure the access success rate for devices of different priorities. As a drawback, as all devices share the same random access channel (RACH) resources, they have the same collision rate. Despite the justifiability of access delay, the system cannot decouple different ACs from each other, so that when mMTCDs generate impulsive RA request bursts, devices of all other ACs will be seriously affected. Due to the same reason, ACB methods cannot grant  any certain AC with a certain access success rate, which may be required for some duty-circle-critical applications such as ultra-reliable communication (URC) in vehicle-to-anything (V2X) communications. Moreover, the amount of ACs is generally fixed to a small value in standards. For instance, 15 ACs in total are defined in LTE-A, including 10 ordinary classes and 5 special ones. This limits its application in mMTC, where a large number of device types can be needed, due to the miscellaneous potential use cases.

\begin{figure}[!h]
	\centering
	\includegraphics[width=.36\textwidth]{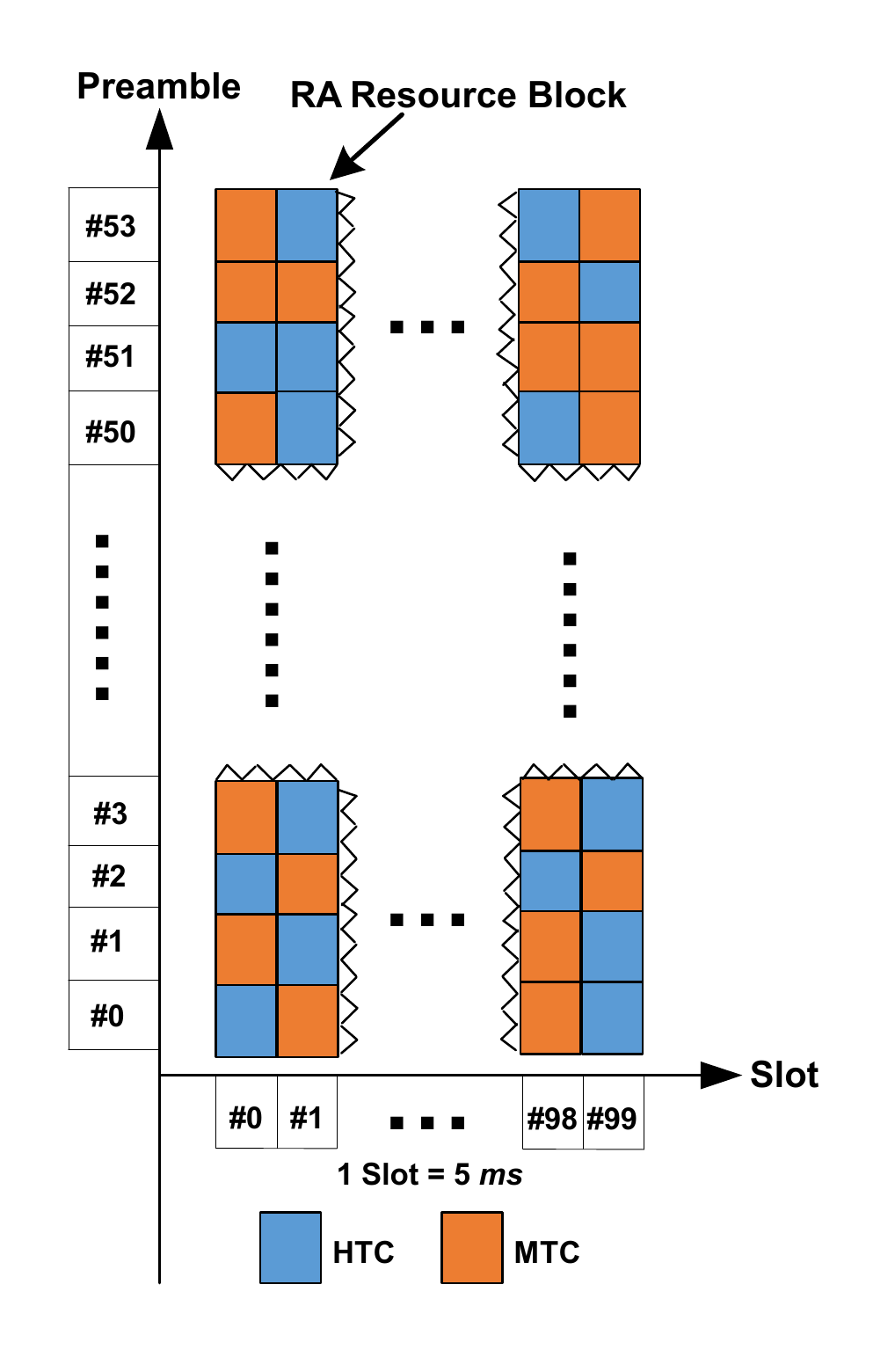}
	\caption{Resource Separation for LTE RA collision control}
	\label{fig:resource_separation}
\end{figure}

In contrast, the RA Resource Separation approach is able to serve devices of various types with individual collision rates. The principle, as illustrated in Fig. \ref{fig:resource_separation} is to separate the available RACH resources into different resource pools, and each pool is dedicated to one certain device type, e.g. Human-Type Communication (HTC) or MTC, so that their collision rates are decoupled from each other. Besides, the implementation of this mechanism is also simple. However, as the RA request density is not decreased, it brings no gain in the average access delay, nor improvement in the average access success rate, while the energy efficiency even drops. To mitigate these drawbacks, it has been proposed to combine this method with the ACB mechanism, known as the prioritized RA\cite{lo2011enhanced}.

%Massive number of connections and huge amount of MTC Devices (MTCDs) can also lead to various technical challenges on the road to 5G networks. The most important challenge among them is the congestion of Radio Access Network (RAN) due to Random Access (RA) collisions. When the number of MTCDs increases, multiple devices activate the same preamble in one Random Access Opportunity (RAO), which furthermore leads to collisions of the Radio Resource Control (RRC) connection request. Most of the MTCDs are unable to initiate connection in the first attempt and therefore perform subsequent attempts due to the high load which also result in the collisions. In order to put a solution to cope with the congestion, the Extended Access Class Barring (EAB) was proposed.  In the EAB, certain classes of MTCDs are temporally blocked from participating in the Access Reservation Protocol (ARP). This scheme leads to two drawbacks, the first one is the cost of an increased access latency to those same MTCDs and the second one is that the network learns the devices' identities and connection establishment causes only after the RRC Connection Request is successfully received. 

In recent years, an emerging category of RA collision controlling methods have been studied, where MTCDs are clustered into different groups\cite{jang2014spatial, lien2011toward}. Here we refer to them as grouped RA.  By reusing a same preamble for all devices in one group, the collision rate can be efficiently reduced. Furthermore, each group can be organized into a master-slave subnetwork, by connecting the group members via device-to-device (D2D) links, in order to improve the RA energy efficiency\cite{tu2011energy,wang2013random,ji2017applying}. It has been proposed, that the clustering process in grouped RA can be combined with a device classification, so that each group only consists of devices of the same class. Upon the RA collision level, RACH resources i.e. Random Access Opportunities (RAOs) can be dedicated to different device classes (DCs), or shared by multiple classes\cite{chatzikokolakis2015way,han2017d2d}. This is an extended version of the RA Resource Separation mechanism in the framework of grouped RA, where multiple classes are used instead of two categories (HTC/MTC) to label devices. As D2D-based grouped RA effectively reduces the RA request density, it well complements the resource separation by nature. Preliminary analysis on the collision rate in this mechanism has been reported in \cite{han2017group}, but the optimization of RACH resource allocation has not been discussed yet.

In this work, we tend to analyze the available RACH resource allocation strategies in grouped RA with respect to the collision rate and collision density. Based on the analysis, we propose two optimized solusions of resource allocation, including an non-prioritized specification which minimizes the overall collision rate, and an adaptive access mechanism which realizes a device class preference without applying ACB.

The rest of the paper is organized as follows. In Sec.\ref{sec:grouped_allocation} we briefly summarize the RACH resource allocation in grouped RA. Subsequently we analyze the three different allocation strategies in Sec.\ref{sec:strategies}, and optimize one of them in Sec.\ref{sec:optimization}. After verifying our result through numerical simulations in Sec.\ref{sec:simulations}, we will conclude this work with Sec.\ref{sec:conclusion}.

\section{Resource Allocation in\\Grouped Random Access}\label{sec:grouped_allocation}
\begin{figure}[!h]
	\centering
	\includegraphics[width=.48\textwidth]{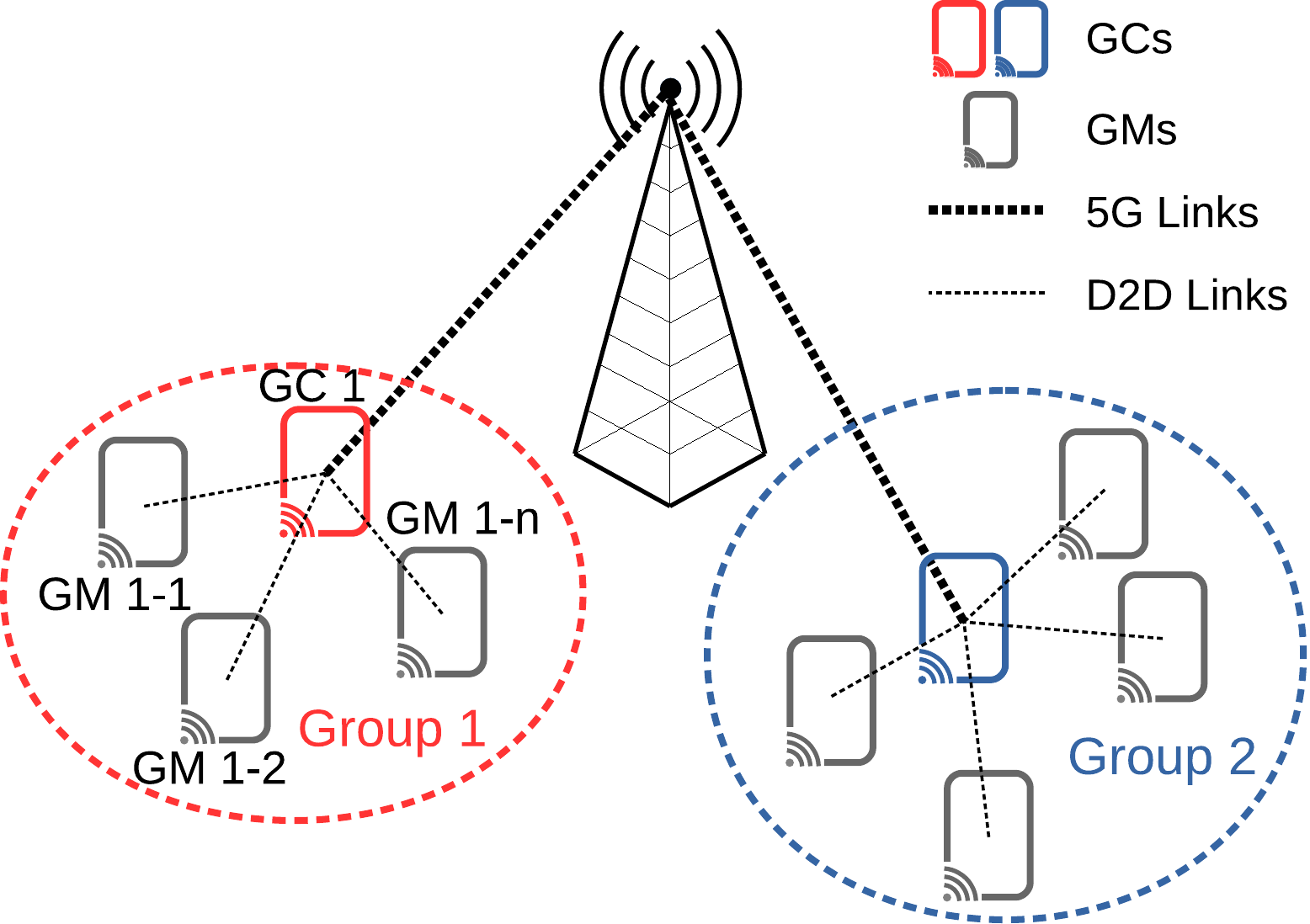}
	\caption{Network topology in grouped RA systems}
	\label{fig:grouped_linkage}
\end{figure}

Our study begins with a review on the D2D-based grouped RA. Methods of this kind have been described in \cite{chatzikokolakis2015way,han2017d2d,han2017group}, and their principle can be illustrated by Fig. \ref{fig:grouped_linkage}. User elements are clustered into groups, every group has one group coordinator (GC), which accesses base-station for the entire group, aggregates uplink data from its group members (GMs) and distributes donwlink data to them over D2D links. Thus, only one device in each group (the GC) needs to request for RA, and the intra-group data aggregation and distribution can be performed between two RA processes. To simplify the timing and intra-group controlling, each group is supposed to consist only devices of the same DC, while multiple groups can be of the same DC. The DC is determined according to device context information such as synchronization mode, periodicity, message delay, application type, etc.

As proposed in \cite{chatzikokolakis2015way}, RACH resources can be flexibly allocated to different DCs. In \cite{han2017group} we have concluded three resource allocation strategies, including:
\begin{itemize}
	\item full sharing: where every RAO can be utilized by all devices;
	\item full dedication: where every RAO is dedicated to devices of a certain DC;
	\item partial dedication: where every RAO can be dedicated to one or more DCs, or available for all devices.
\end{itemize}

\section{Collision under Different Resource Allocation Strategies}\label{sec:strategies}
To analyze the impact of different resource allocation strategies on the collision rate in grouped RA, first we review the traditional simple RA case, where neither device classification nor resource separation is applied. Assume that all the devices in local cell generate $\gamma$ RA requests per second, and $L$ RAOs in total are available in each second. According to \cite{3gpp2011study}, when $\gamma$ is large, the probability of access success for each attempt can be approximately calculated by
\begin{equation}
	q = e^{-\frac{\gamma}{L}},
\end{equation}
and hence the collision rate
\begin{equation}
	p = 1-q = 1-e^{-\frac{\gamma}{L}}.
\end{equation}

Then, we extend this collision rate model to the case of grouped RA with device classification, under different resource allocation strategies. We assume that devices in the local cell are categorized into $N$ different DCs, and the devices of each DC $i\in\{1,2,\dots,N\}$ generate $\gamma_i$ RA requests in total per second. Obviously, we have
\begin{equation}
	\gamma=\sum\limits_{i=1}^N\gamma_i.
\end{equation}

\subsection{Full Sharing}
The case of full sharing strategy is simple and straight forward. As every device is able to randomly choose one RAO from all the $L$ options to send its RA request, it can cause a collision with any other device, no matter if they are of the same DC. In this case, the collision rate can be computed as
\begin{equation}
	p_\textrm{FS}=1-e^{-\sum\limits_i\frac{\gamma_i}{L}}=1-e^{-\frac{\gamma}{L}}=p,
\end{equation}
which is same as the simple RA case without device classification.

\subsection{Full Dedication}
Then we investigate the full dedication case. Assuming that for each DC $i$, $L_i$ RAOs are dedicated, and that all $L$ RAOs are allocated. Due to the resource separation, no collision can occur between two devices of different DCs. Hence the collision rate for the DC $i$ can be computed as
\begin{equation}
p_{\textrm{FD},i}=1-e^{-\frac{\gamma_i}{L_i}},\label{equ:collision_rate_fd}
\end{equation}
where 
\begin{equation}
	L=\sum\limits_{i=1}^NL_i.
\end{equation}

\subsection{Partial Dedication}
The case of partial dedication is the most complex one. For each DC $i$, we denote the set of RAOs it can deploy (including dedicated and shared) as $\mathbb{B}_i$.
First, for an arbitrary device of DC $i$, when it attempts to send a RA request at the RAO $l$, the collision rate is
\begin{equation}
	p_{\textrm{PD},i,l}=1-e^{-\sum\limits_{j\in\mathbb{A}_l}\frac{\gamma_j}{L_j}},\label{equ:collision_rate_pd_1}
\end{equation}
where $\mathbb{A}_l$ denotes the set of all DCs which share the RAO $l$. Next, as the device randomly selects one RAO from $\mathbb{B}_i$ for this attempt, the collision rate for devices of class $i$ in each RA attempt is
\begin{equation}
	p_{\textrm{PD},i}=\frac{1}{\sharp\mathbb{B}_i}\sum\limits_{l\in\mathbb{B}_i}p_{\textrm{PD},i,l}.\label{equ:collision_rate_pd_2},
\end{equation}
where $\sharp\mathbb{B}_i$ represents the amount of elements in set $\mathbb{B}_i$.

\section{Optimized Random Access Channel\\Resource Dedication}\label{sec:optimization}
Generally, the full sharing strategy is inflexible, and exhibits a fixed performance depending only on the overall RA request density and RAO amount. No dynamic optimization can be applied under this strategy. In contrast, the partial dedication policy provides a high degree of freedom in RAO allocation, but the collision rate cannot be minimized with a reasonable computation cost due to its complex mathematical form presented in (\ref{equ:collision_rate_pd_1}--\ref{equ:collision_rate_pd_2}). Therefore, it is not attractive for optimization either. Under the full dedication strategy, however, the collision rate $p_{\textrm{FD},i}$ is a simple function of $L_i$, providing an enough potential of RACH resource optimization with cost efficiency. Hence, in the rest part of this paper, we focus on this strategy for our discussion about the RACH resource optimization .

\subsection{Approximation: Independent Collisions}
Despite of the simple form (\ref{equ:collision_rate_fd}) of the collision rate for each single RA attempt, it is still challenging to accurately model the overall collision density in the local cell, due to the correlation between different collision events, as mentioned in Sec. \ref{sec:strategies}. To simplify the computation, we take an approximate assumption, that the RA collision processes of different accessing devices are independent to each other. Under this assumption and the full RAO dedication strategy, the overall collision density in local cell can be easily calculated as
\begin{equation}
C_{\textrm{FD,cell}}=\sum\limits_{i=1}^N\gamma_ip_{\textrm{FD},i},\label{equ:approx_density}
\end{equation}
and the overall probability of collision occurrence is
\begin{equation}
p_\textrm{FD,cell}=1-\prod\limits_{i=1}^N(e^{-\frac{\gamma_i}{L_i}})^{\gamma_i}=1-\prod\limits_{i=1}^Ne^{-\frac{2\gamma_i}{L_i}}\label{equ:equ:approx_collision_rate}
\end{equation}

\subsection{Overall Collision Rate Minimization}
To implement an global optimization without DC preference. we attempt to minimize $p_\textrm{FD,cell}$.
Given certain values $\gamma_i$ for all $i\in\{1,2,\dots,N\}$, there is always 
\begin{equation}
	e^{-\frac{2\gamma_i}{L_i}}>0.%\quad\forall i\in\{1,2,\dots,N\},
\end{equation}
Thus, according to the AM-GM inequality \cite{steele2004cauchy} there is
\begin{equation}
	 p_\textrm{FD,cell}\ge 1-(\frac{1}{N}\sum\limits_{i=1}^Ne^{-\frac{2\gamma_i}{L_i}})^N.
\end{equation}
The equality holds if and only if
\begin{equation}
	\frac{\gamma_i}{L_i}=\frac{\gamma_j}{L_j}\quad[i,j]\in\{1,2,\dots,N\}^2,\label{equ:optimal_allocation}
\end{equation}
which minimizes $p_\textrm{FD,cell}$.

\subsection{Random Access with Class Preference}
In practice, some applications may have higher priorities to be granted with RACH resources than others, in order to fulfill their special requirements in quality of service (QoS). For instance, emergency services and public utilities should be guaranteed with better access chances, like LTE-A systems provide them with the special access classes 13 and 14. Here, we investigate the essential amount of RACH resource for a certain QoS requirement such as expected collision rate and average access delay.

\subsubsection{Collision-Rate-Oriented RAO Reservation}
To achieve a certain requirement of collision rate $\hat{p}_i$ for the device class $i$, according to (\ref{equ:collision_rate_fd}), under the full dedication strategy there shall be
\begin{equation}
	L_i\geq\frac{\gamma_i}{-\ln(1-\hat{p}_i)}\label{equ:collision_rate_reservation}
\end{equation}

\subsubsection{Access-Delay-Oriented RAO Reservation}
Under the full dedication strategy, the average access delay of device class $i$ can be calculated by
\begin{equation}
\begin{split}
	D_i =&T_i(1-p_\textrm{FD,i})p_\textrm{FD,i}+2T_i(1-p_\textrm{FD,i})p_\textrm{FD,i}^2\\
    &+3T_i(1-p_\textrm{FD,i})p_\textrm{FD,i}^3+\dots\\
    =&T_i(1-p_\textrm{FD,i})\sum\limits_{n=1}^\infty np_\textrm{FD,i}^n,
\end{split}
\end{equation}
where $T_i$ is the time-off before reattempting RA for each device of class $i$. Note that $np_\textrm{FD,i}^n$ is an arithmetico-geometric series, whose sum to the infinite term can be computed according to \cite{dinesh2010pearson} as
\begin{equation}
	\sum\limits_{n=1}^\infty np_\textrm{FD,i}^n=\frac{1}{1-p_\textrm{FD,i}}+\frac{p_\textrm{FD,i}}{(1-p_\textrm{FD,i})^2}.
\end{equation}
Hence we have
\begin{equation}
	D_i=\frac{T_i}{1-p_\textrm{FD,i}}=T_ie^{\frac{\gamma_i}{L_i}}
\end{equation}
Therefore, given a certain requirement of average access delay $\bar{D_i}$, it should be fulfilled that
\begin{equation}
	L_i\geq\frac{\gamma_i}{\ln D_i-\ln T_i}.\label{equ:access_delay_reservation}
\end{equation}
Note that this is equivalent to the collision-rate-oriented reservation (\ref{equ:collision_rate_reservation}), because $D_i$ is a function of $\hat{p}_i$ and $T_i$.

\subsubsection{Reserve-and-Divide Method}
To realize a preference for special DCs in the optimized RACH resource allocation, we designed a \textit{Reserve-and-Divide} method. First, essential resources are reserved for the special DCs with respect to their QoS requirements. Afterwards, the unreserved resources are divided by the devices of normal classes, according to the criteria of overall collision rate minimization. If the entire RACH resource pool cannot fulfill the requirements of all special DCs, an exception is reported to trigger extra solutions, such as re-clustering the devices to reduce the accessing devices, or to dynamically reconfigure the network for more RACH resources.

For instance, we define the $1^\textrm{st}$ to the $M^\textrm{th}$ out of the $N$ DCs as special classes. Given a collision-rate requirement for each of them, denoted as $\hat{p}_i$ where
$i=1,2,\dots,M$, we can implement the reserve-and-allocate method as briefly presented in Fig. \ref{fig:reserve_and_allocate}.

%\begin{figure}[!h]
%	\begin{algorithmic}[1]
%%	\Procedure{Allocate}{$L,N,\gamma_1,\gamma_2,\dots,\gamma_N,\hat{p}_1,\hat{p}_2,\dots,\hat{p}_M$}
%	\For{$i=1\to M$}\Comment{Reserve RAOs for special DCs}
%	\State $L_i\gets\frac{\gamma_i}{-\ln(1-\hat{p}_i)}$
%	\State $L\gets L-L_i$
%	\If{$L<0$}
%	\State \textit{Error: RACH resource overload}
%	\EndIf
%	\EndFor
%	\For{$i=M+1\to N$}\Comment{Allocate the rest RAOs}
%	\State $L_i\gets\frac{L\gamma_i}{\sum_{j=M+1}^N\gamma_j}$
%	\EndFor	
%	\State\Return{$[L_1,L_2,\dots,L_N]$}
%%	\EndProcedure
%	\end{algorithmic}
%	\caption{Algorithm for ...}
%\end{figure}

\begin{figure}[!h]
	\removelatexerror
	%	\begin{algorithmic}[1]
	\begin{algorithm}[H]
		%\caption{Algorithm for ...}
		%	\Procedure{Allocate}{$L,N,\gamma_1,\gamma_2,\dots,\gamma_N,\hat{p}_1,\hat{p}_2,\dots,\hat{p}_M$}
		%	\For {$i=1\to M$}\Comment{Reserve RAOs for special DCs}
		%	\State $L_i\gets\frac{\gamma_i}{-\ln(1-\hat{p}_i)}$
		%	\State $L\gets L-L_i$
		%	\If{$L<0$}
		%	\State \textit{Error: RACH resource overload}
		%	\EndIf
		%	\EndFor
		%	\For{$i=M+1\to N$}\Comment{Allocate the rest RAOs}
		%	\State $L_i\gets\frac{L\gamma_i}{\sum_{j=M+1}^N\gamma_j}$
		%	\EndFor	
		%	\State\Return{$[L_1,L_2,\dots,L_N]$}
		%	\EndProcedure
		%	\end{algorithmic}
		initialize $L,N,\gamma_1,\gamma_2,\dots,\gamma_N,\hat{p}_1,\hat{p}_2,\dots,\hat{p}_M$\;
		\For( \emph{Reserve RAOs for special DCs}){$i = 1$ to $M$}
		{
			$L_i\gets-\gamma_i/\ln(1-\hat{p}_i)$\;
			$L\gets L-L_i$\;
			\If{$L<0$}{Exception: RACH resource overload}
		}
		\For( \emph{Allocate the rest RAOs}){$i = M+1$ to $N$}
		{
			$L_i\gets L\gamma_i/\sum_{j=M+1}^N\gamma_j$\;
		}
	\end{algorithm}
	\caption{The proposed \textit{Reserve-and-Divide} algorithm with collision rate requirements provided}
	\label{fig:reserve_and_allocate}
\end{figure}
  
\section{Simulations}\label{sec:simulations}
\subsection{Overall Collision Rate Minimization}
To briefly verify the feasibility of our proposed optimizing method (\ref{equ:optimal_allocation}), which bases on the approximated assumption of independent collisions (\ref{equ:approx_density}--\ref{equ:equ:approx_collision_rate}), we executed numerical simulations to test the optimal RACH resource allocation between two DCs. We defined four different reference DCs, as listed in Tab. \ref{table:dc_config_sim1}, and ran the simulation for three different configurations of DC coexistence: DCs 1+2, DCs 1+3 and DCs 1+4. Like the current LTE-A specification, we assumed that in every second there are $54\times200=10 800$ RAOs available. For each simulation, 500 iterations of Monte-Carlo test were executed to reduce the deviation.
\begin{table}[!h]
	\caption{Device class specifications. For simplification, all devices are assumed to access the network randomly and asynchronously.}
	\label{table:dc_config_sim1}
	\centering
	\begin{tabular}{c|c|c|c}
		\textbf{Class} & \textbf{Accessing devices} & \textbf{Avg. access frequency}  & \textbf{RA density}\\\hline
		1 & 3000 & \SI{1/60}{\hertz} & \SI{50}{\hertz}\\\hline
		2 & 30 000 & \SI{1/300}{\hertz} & \SI{100}{\hertz}\\\hline
		3 & 30 000 & \SI{1/60}{\hertz} & \SI{500}{\hertz}\\\hline
		4 & 30 000 & \SI{1/30}{\hertz} & \SI{1000}{\hertz}
	\end{tabular}
\end{table}

\begin{figure}[!h]
	\centering
	\includegraphics[width=.48\textwidth]{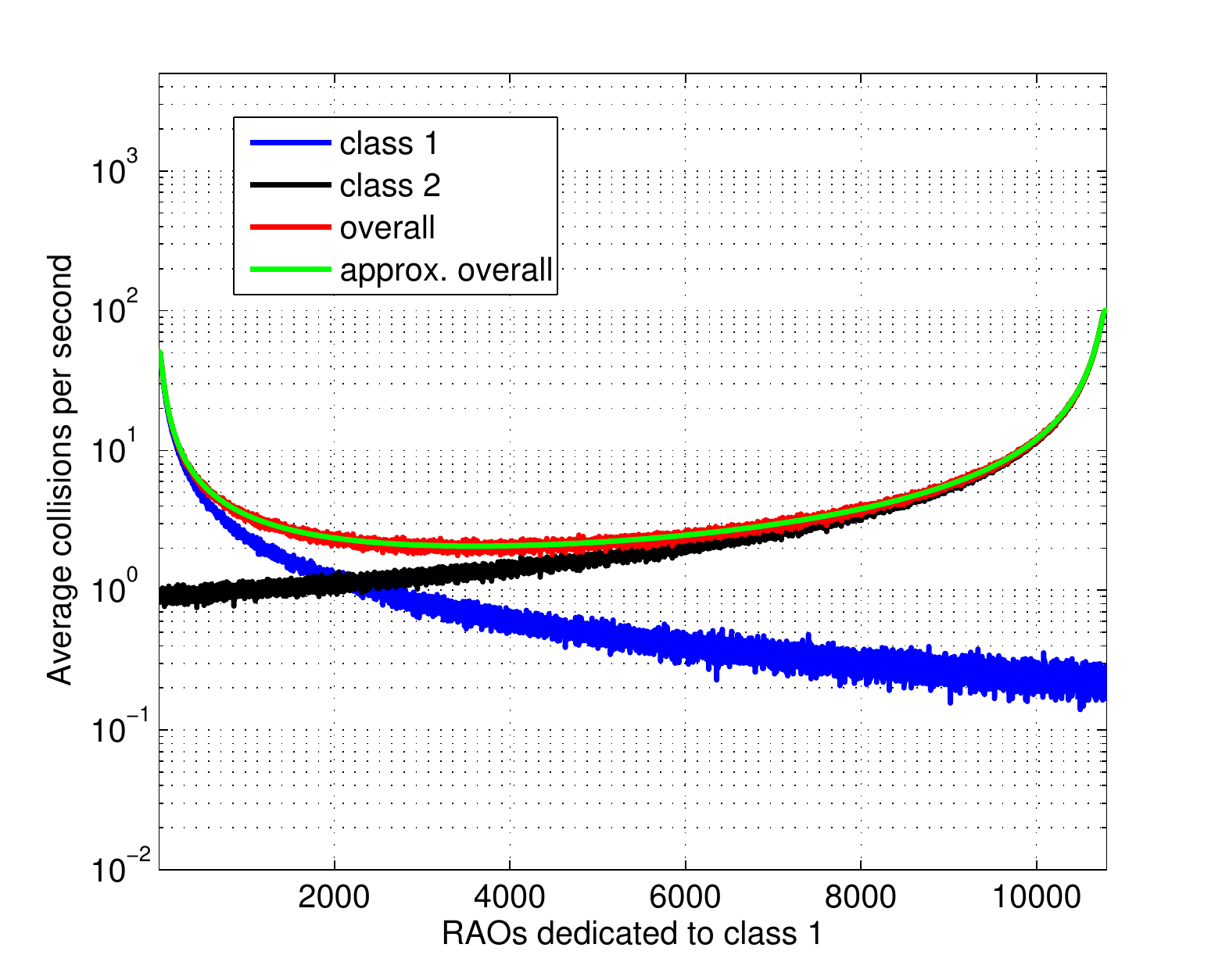}
	\caption{Collision densities with respect to RACH resource dedication to DC1 and DC2. Numerical results obtained from 500 iterations of Monte-Carlo test.}
	\label{fig:opt_allocation_DC1+DC2}
\end{figure}

Take the first configuration as instance, the result is shown in Fig. \ref{fig:opt_allocation_DC1+DC2}. Clearly, the density of collisions generated by devices of DC1 decreases with more RAOs dedicated to DC1, while the DC2-generated collisions is increasing. A minimum of the overall collision density can be achieved, as the red curve shows. The approximated theoretical model (green curve) well matches the simulation result, and gives a reasonable estimation of optimal $L_1$. 

To evaluate the accuracy of (\ref{equ:optimal_allocation}), we took the estimated optimal $L_1$, and investigated the collision density under this allocation according to the simulation. Subsequently we compare it to the minimal collision density obtained through the simulation. We also tested the full sharing strategy for reference. This evaluation was carried out for all three DC combinations, and the results are listed in Tab. \ref{table:comp_est_sim_allocation}. It can be obtained that the estimator generally returns satisfying results, and the error decreases with rising collision density. Furthermore, the optimal full dedication leads to a performance similar to the full sharing strategy. However, taking the fact into account, that some device classes can exhibit highly time-varying RA request density, the full dedication strategy surpasses full sharing, as it isolates different DCs from each other, so that RA bursts generated by a special DC will not impact the access chance of other DCs.

\begin{table}[!h]
	\caption{Comparing estimated and simulated optimum of RACH resource allocation}
	\label{table:comp_est_sim_allocation}
	\centering
	\begin{tabular}{c|c|c|c}
		\textbf{DC Combination} 															 & 1\&2 			& 1\&3 			& 1\&4  \\\hline
		\textbf{Optimal. $L_1$ (estimated)} 						 					& 3600 		  	& 982  			  & 514\\\hline
		\textbf{Optimal. $L_1$ (simulated)} 					 	  					& 3460 		   	& 1048 			 & 534\\\hline
		\textbf{Collision density at est. opt. $L_1$ (\SI{}{\hertz})} 		& 1.766 		& 26.320 		& 95.690\\\hline
		\textbf{Collision density at sim. opt. $L_1$ (\SI{}{\hertz})} 		& 1.998 		& 26.780 		& 97.042\\\hline
		\textbf{Collision density under full sharing (\SI{}{\hertz})}			& 2.166		& 26.940		& 97.634
	\end{tabular}
\end{table}

\subsection{Reserve-and-Divide Method}
\begin{figure}[!h]
	\centering
	\includegraphics[width=.48\textwidth]{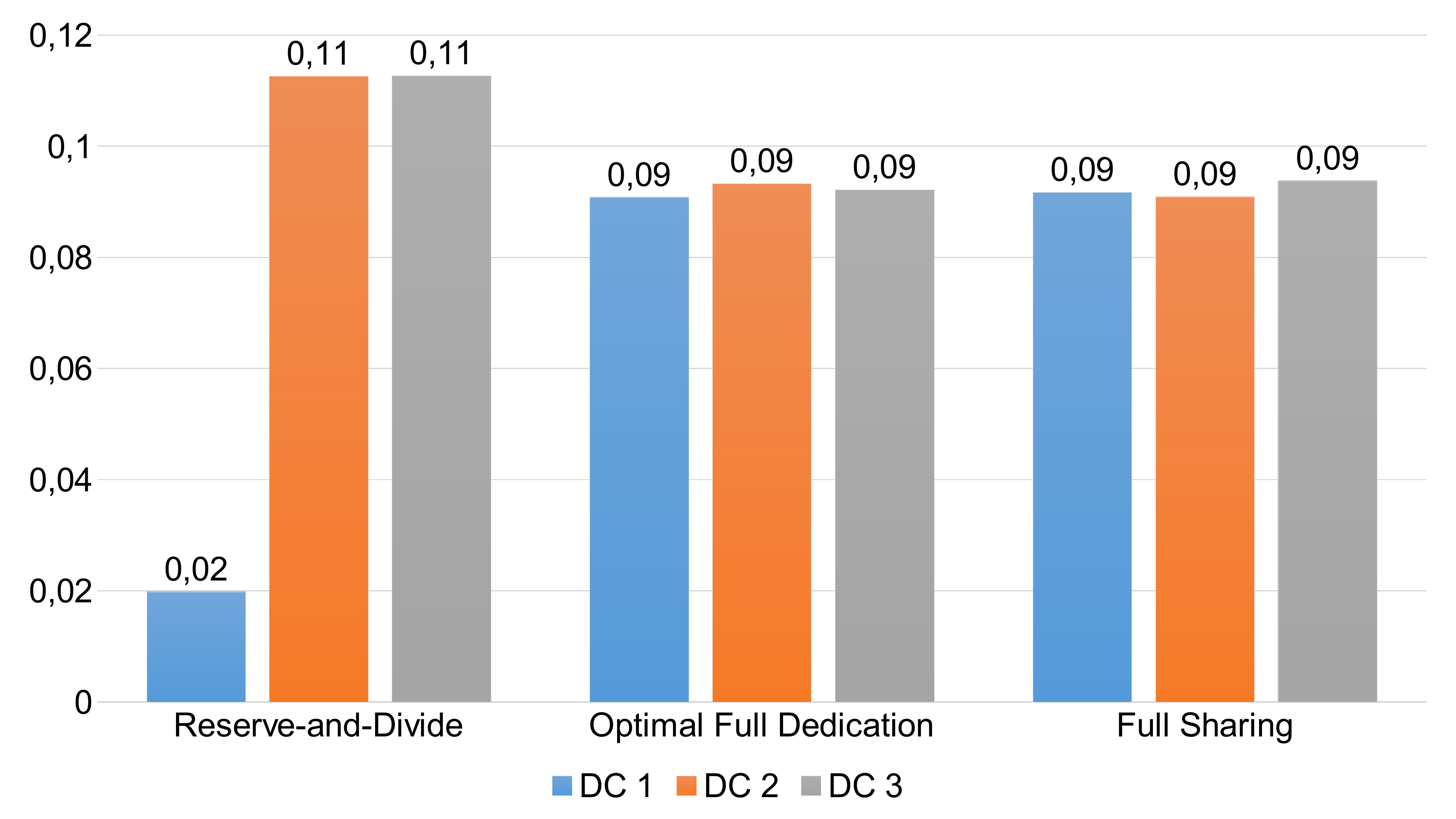}
	\caption{Collision rates in 3 device classes under different resource allocation approaches.}
	\label{fig:reserve_and_divide}
\end{figure}

As the next step, we attempted to verify the proposed reserve-and-divide method through simulations, where the DCs 1 to 3 coexist in the local cell, and a maximal collision rate of $0.02$ is required for DC1. 100 000 iterations of Monte-Carlo tests were carried out to obtain the achieved collision rates under the \textit{Reserve-and-Divide}, the optimal full dedication and the full sharing strategies, as compared in Fig. \ref{fig:reserve_and_divide}. It can be seen that the proposed method can efficiently guarantee the special requirement of DC1. 

\section{Conclusion}\label{sec:conclusion}
So far, we have analyzed the RA collision rate and the RA collision density in grouped RA systems under different RACH resource allocation strategies. Focusing on the full dedication strategy, we have proposed an approximate estimator of optimal resource allocation, and an additional method to realize device class preference. Numerical simulations have verified the feasibility of our proposed methods.

For future work, the methods should be tested under complex scenarios with more coexisting device classes for a better verification. Dynamic update of the resource allocation should be investigated to deal with the highly time-varying RA density in mMTC applications. Furthermore, the impact of the proposed methods on the device battery life shall be investigated, and the potential of integrating them with further power efficient technologies such as \cite{schulist2010requesting} should be discussed.


\section*{Acknowledgment}
This work has been performed with supports of the H2020-ICT-2014-2 project 5G NORMA and the H2020-MSCA-ITN-2015 project 5Gaura. The authors would like to acknowledge the contributions of their colleagues. This information reflects the consortia's view, but the consortia are not liable for any use that may be made of any of the information contained therein.

\bibliographystyle{IEEEtran}
\bibliography{references}

\end{document}